\documentclass[12pt]{article}
\usepackage[a4paper, total={6.2in, 9in}]{geometry}
\usepackage[nodayofweek]{datetime}
\usepackage{multirow}
\usepackage{physics}
\usepackage{graphicx}
\usepackage{epstopdf}
\usepackage{booktabs}
\PassOptionsToPackage{hyphens}{url}\usepackage{url}
\usepackage[hidelinks]{hyperref}
\hypersetup{colorlinks,citecolor=blue}
\usepackage{fancyhdr}                           
\usepackage[bottom]{footmisc}
\usepackage{lmodern}
\usepackage{xcolor}
\hypersetup{colorlinks = false
}

\hypersetup
{
    pdfauthor={Dimitrios Emmanoulopoulos},
    pdfsubject={Research Paper for Barclays Bank PLC.},
    pdftitle={Quantum Machine Learning in Finance: Time Series Forecasting},
    pdfkeywords={Quantum Computing, Quantum Machine Learning, Neural Networks, Machine Learning, biLSTM, PQC, bidirectional long short-term memory, parametrised quantum circuits }
}
\newdate{date}{01}{02}{2022}
\date{\displaydate{date}}

\fancypagestyle{firstpage} 
{
   \fancyhf{}
   \fancyfoot[C]{\vspace{-12mm}\thepage}

   \fancyfoot[L]{\vspace{-5mm}\noindent\rule{6.2in}{0.4pt}
     \scriptsize{
       \textcopyright \hspace{0.5mm} Barclays Bank PLC 2022 \\
       This work is licensed under a \href{https://creativecommons.org/licenses/by/4.0/}
       {Creative Commons Attribution 4.0 International License (CC BY).}  \\ 
       Provided you adhere to the CC BY license, including as to attribution, you are
       free to copy and redistribute this work in any medium or format and remix, transform,
       and build upon the work for any purpose, even commercially.                  \\
       BARCLAYS is a registered trade mark of Barclays Bank PLC, all rights are reserved. \\
       $^\star$Sofija Dimoska is no longer with Barclays.\\
     }
   }
}
\title{\vspace{-2cm}\hspace{1.5em}Quantum Machine Learning in Finance:\newline \hspace*{-0.5em}Time Series Forecasting}

\author{%
  \begin{tabular}{c} {\fontsize{10.75}{1cm}\selectfont Dimitrios Emmanoulopoulos and Sofija Dimoska$^\star$} \\ {\fontsize{10.75}{1cm}\selectfont Chief Technology Office} \\
    {\fontsize{10.75}{1cm}\selectfont Barclays} \\ \hskip 1em \end{tabular} }

\date{\vspace{-0.5cm}February 1, 2022}

\begin{document}
\maketitle 
\thispagestyle{firstpage}
\begin{abstract}
\noindent We explore the efficacy of the novel use of parametrised quantum circuits (PQCs) as quantum neural networks (QNNs) for forecasting time series signals with simulated quantum forward propagation. The temporal signals consist of several sinusoidal components (deterministic signal), blended together with trends and additive noise. The performance of the PQCs is compared against that of classical bidirectional long short-term memory (BiLSTM) neural networks. Our results show that for time series signals consisting of small amplitude noise variations (up to 40 per cent of the amplitude of the deterministic signal) PQCs, with only a few parameters, perform similar to classical BiLSTM networks, with thousands of parameters, and outperform them for signals with higher amplitude noise variations. Thus, QNNs can be used effectively to model time series having, at the same time, the significant advantage of being trained significantly faster than a classical machine learning model in a quantum computer. 
\end{abstract}

\section{Introduction}
\label{sect:intro}

In a genuine quantum computing infrastructure (i.e.~an environment which performs computations using quantum phenomena such as superposition, entanglement, tunnelling etc.) quantum algorithms \cite{QZoo} can run significantly faster with respect to their classical equivalents. These algorithms cover many application areas such as optimisation problems, combinatorics, cryptography, solution of partial differential equations, simulations e.g.~\cite{montanaro2016quantum}.

In the field of machine learning quantum enhancement can come in two different flavours e.g.~\cite{huang2021power}. The first flavour concerns is around the training process which can be significantly accelerated, even for classical machine learning models, using quantum optimisation algorithms. In some cases, involving complex loss functions with multiple minima, these algorithms can even yield more accurate results by localising more effectively the global minimum. The second flavour concerns the identification of complex patterns since, in a quantum computing environment, the data as they can be sampled more efficiently from probability distributions that are exponentially difficult to sample using classical methods on classical computers.

The first generation of quantum machine learning (QML) focused on the training process which was accelerated by quantum algorithms \cite{QML}, yielding a plethora of quantum equivalents of classical machine learning methods e.g.~Quantum Boltzmann Machine \cite{wiebeBolzman}, Quantum PCA \cite{QPCA}, Quantum SVM \cite{QSVM}. After the advent of Noisy Intermediate-Scale Quantum processors \cite{Preskill2018quantumcomputingin} the field of QML was evolved more towards deep neural networks \cite{killoran2019continuous,beer2020training,abbas2021power,bauschRQNN2020} known as Quantum Neural Networks (QNNs). The majority of these deep neural network algorithms use Parametrised Quantum Circuits (PQCs)\cite{verdon2017quantum,mitarai2018PQC} and this term is now used equivalently with the term QNNs \cite{benedetti2019parameterized}. PQCs can be designed to describe patterns that relate the quantum encoded classical input to the state of a readout \cite{fahriClassification18}. Application of PQCs \cite{2020BroughtonQML}
 to the MNIST database sample \cite{lecun1998mnist} has been shown to have a good accuracy of around 90 per cent similar to that of an equivalent (i.e.~ with respect to the number of parameters) classical neural network.

Classical neural networks have a vast variety of applications \cite{liu2017survey} e.g.~computer vision, speech and language applications, drug design, anomaly detection, recommendation systems. Recurrent neural networks \cite{pineda1987generalization,pearlmutter1989learning} are a class of neural networks that are used very successfully to model sequential data e.g.~speech recognition \cite{graves13RNNspeech}, music composition \cite{eckRNNmusic02}, handwriting recognition \cite{graves2008offline}. Time series data are by default sequential data and thus RNNs have been used extensively in order to model them \cite{HEWAMALAGE2021388,petnehazi2019recurrent}.

Particularly for time series forecasting, the RNN architecture of long short-term memory (LSTM) neural networks have been proven very efficient predictors to a plethora of use cases e.g.~ weather prediction \cite{zaytar2016sequence}, traffic flow predictions \cite{rui16TF}, biological sequence data analysis \cite{sonderby2015convolutional}. Moreover, bidirectional LSTM (BiLSTM) neural networks \cite{graves2005framewise} are able to disclose even more information from a given temporal sequence \cite{SiamiNaminiLSTM19}, improving even more the context available to the algorithm. They consist of two LSTMs, one taking input in a forward direction, and the other in a backwards direction and their power relies on the fact of knowing what outputs follow and precede immediately after and before, respectively, of a given input in a signal. For quantum computing, there are quantum equivalents of an RNN and an LSTM neural network, the quantum RNN \cite{bauschRQNN2020} and quantum LSTM \cite{QLSTM}, respectively.

Time series in the financial sector play a major role in many business areas for pricing, asset management, quant strategies, and risk management. LSTM and BiLSTM neural networks have been used very successfully to model these types of signals \cite{selvin2017stock,nelson2017stock,kim2018forecasting,li2018stock,jin2020stock,sunny2020deep} and their forecasting power has been proven to be superior from that of the traditional auto-regressive type statistical methods \cite{heaton17,siami2018comparison,Kandappan21}. 

In this work we compare the predictive power of QNNs, represented by a the PQC, to that of a BiLSTM neural network initially for periodic signals with additive noise and trends and then for more complex temporal signals. In Section~\ref{sect:background} we give a short background information around quantum computing and QNNs and in Section~\ref{sect:simulations} we describe in detail the entire simulation process, including the temporal signal creation, the data preprocessing, the quantum data encoding and the structure of both the quantum and classical neural networks. In Section~\ref{sect:comparison} we compare our results using statistical metrics as derived from the test (i.e.~out of time) temporal data and in Section~\ref{sect:summ_discu} we discuss our findings together with some future work. 

All the experiments are conducted using \texttt{Python\;3.6}. For the BiLSTMs we are using \texttt{Keras\;2.4.3}\footnote{\url{https://github.com/keras-team/keras}} trained on the top of \texttt{TensorFlow\;2.3.0} \cite{abadi2016tensorflow}. For the design of the PQCs we are using \texttt{Google Cirq\;0.10.0}\footnote{\url{https://quantumai.google/cirq}}  which are then transformed into tensors using \texttt{TensorFlow Quantum\;0.4.0} \cite{2020BroughtonQML}. 

\section{Overview and background information}
\label{sect:background}
In this section we give a brief background to the various concepts discussed throughout the paper regarding quantum computing and parametrised quantum circuits which in these context correspond to QNNs. This is meant to give merely some references points for the readers who are not familiar with these topics, and should not be considered as a review of the quantum computing field.

\subsection{Quantum computing}
\label{sect:QC}
In classical computers, information is stored in bits (a bit is the base logical value which can be either 0 or 1) which are processed with a finite set of digital gates \cite{steane1998quantum}. In quantum computers, the information is stored in quantum bits, called qubits, that can exist in any quantum superposition of two physically independent quantum states, 0 and 1. When a qubit's value is read, it can only be 0 or 1 as the wave function describing the quantum system collapses. Any two-level quantum mechanical system can potentially be used as a qubit \cite{jazaeri2019review} e.g.~ photons, atoms, electrons, nuclei, Josephson junctions, quantum dots etc. Quantum mechanical systems exhibit quantum phenomena such as superposition, entanglement and tunnelling which are used by a quantum computer, via the appropriate algorithms, to reduce the execution time of a given computation. 

The most common way to mathematically represent the qubit state is with a column vector in a Hilbert space whose elements correspond to the probabilities of the qubit being measured in a particular state \cite{kaye2007introduction}. Equation~\ref{eq:qubit} shows a representation of qubit $q$ that can be measured as 0 with probability $q_0$ or measured as 1 with probability $q_1=1-q_0$.

\begin{equation}
\ket{q} = 
\begin{bmatrix}
q_0 \\
q_1 \\
\end{bmatrix}=
\begin{bmatrix}
q_0 \\
1-q_0 \\
\end{bmatrix}
\label{eq:qubit}
\end{equation}

While one qubit can be measured in two states, $n$ qubits can be measured in $2^n$ states as a Kronecker product. Equation~\ref{eq:3qs} shows the mathematical representation of three qubits a, b and c, respectively, describing $2^3$ states. 

\begin{equation}
\ket{c b a} = 
\begin{bmatrix}
c_0 \\
c_1 \\
\end{bmatrix}
\otimes
\begin{bmatrix}
b_0 \\
b_1 \\
\end{bmatrix}
\otimes
\begin{bmatrix}
a_0 \\
a_1 \\
\end{bmatrix}
=
\begin{bmatrix}
c_0 \begin{bmatrix} 
b_0 \\
b_1
\end{bmatrix} \\
c_1 \begin{bmatrix}
 b_0 \\
 b_1
\end{bmatrix}\\
\end{bmatrix}
\otimes
\begin{bmatrix}
a_0 \\
a_1 \\
\end{bmatrix}
=
\begin{bmatrix}
c_0b_0
\begin{bmatrix}
a_0 \\
a_1 \\
\end{bmatrix} \\
c_0b_1
\begin{bmatrix}
a_0 \\
a_1 \\
\end{bmatrix} \\
c_1b_0 
\begin{bmatrix}
a_0 \\
a_1 \\
\end{bmatrix}\\
c_1b_1 \begin{bmatrix}
a_0 \\
a_1 \\
\end{bmatrix}
\end{bmatrix}
=
\begin{bmatrix}
c_0b_0a_0\\
c_0b_0a_1\\
c_0b_1a_0\\
c_0b_1a_1\\
c_1b_0a_0\\
c_1b_0a_1\\
c_1b_1a_0\\
c_1b_1a_1\\
\end{bmatrix}
\label{eq:3qs}
\end{equation}
\\

Any change of the qubit states can be modelled mathematically as a multiplication with the corresponding transformation matrix and can be visually modelled as rotations on the Bloch sphere. For example the application of an Hadamard gate, a Pauli-Z gate, and another Hadamard gate to a qubit $q$ can be expressed mathematically in the following way 

\begin{equation}
X \ket{q} = X
\begin{bmatrix}
q_0 \\
q_1 \\
\end{bmatrix}
= \frac{1}{\sqrt{2}}
\begin{bmatrix}
1 & 1 \\
1 & -1\\
\end{bmatrix}
\begin{bmatrix}
1 & 0 \\
0 & -1\\
\end{bmatrix}
\frac{1}{\sqrt{2}}
\begin{bmatrix}
1 & 1 \\
1 & -1\\
\end{bmatrix}
\begin{bmatrix}
q_0 \\
q_1 \\
\end{bmatrix}
=
\begin{bmatrix}
0 & 1 \\
1 & 0\\
\end{bmatrix}
\begin{bmatrix}
q_0 \\
q_1 \\
\end{bmatrix}
=
\begin{bmatrix}
q_1 \\
q_0 \\
\end{bmatrix}
\label{eq:states}
\end{equation}

\begin{figure}
\centering
	\parbox{1.1\linewidth}{
        \includegraphics[height=1.9in, trim={2.5cm 0 0 0},clip]{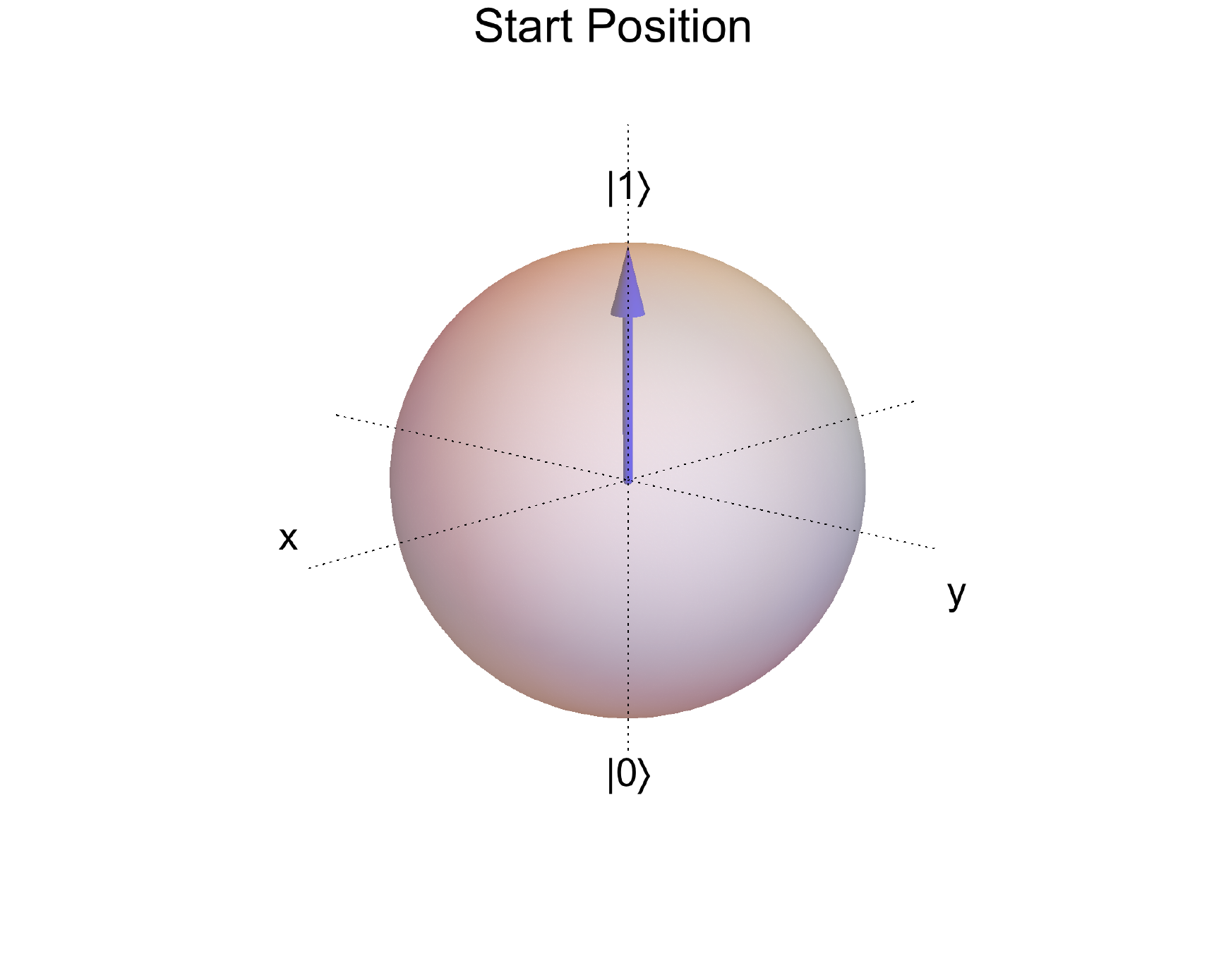}
	\hspace{-4.5 em}
	\includegraphics[height=1.9in, trim={2.5cm 0 0 0},clip]{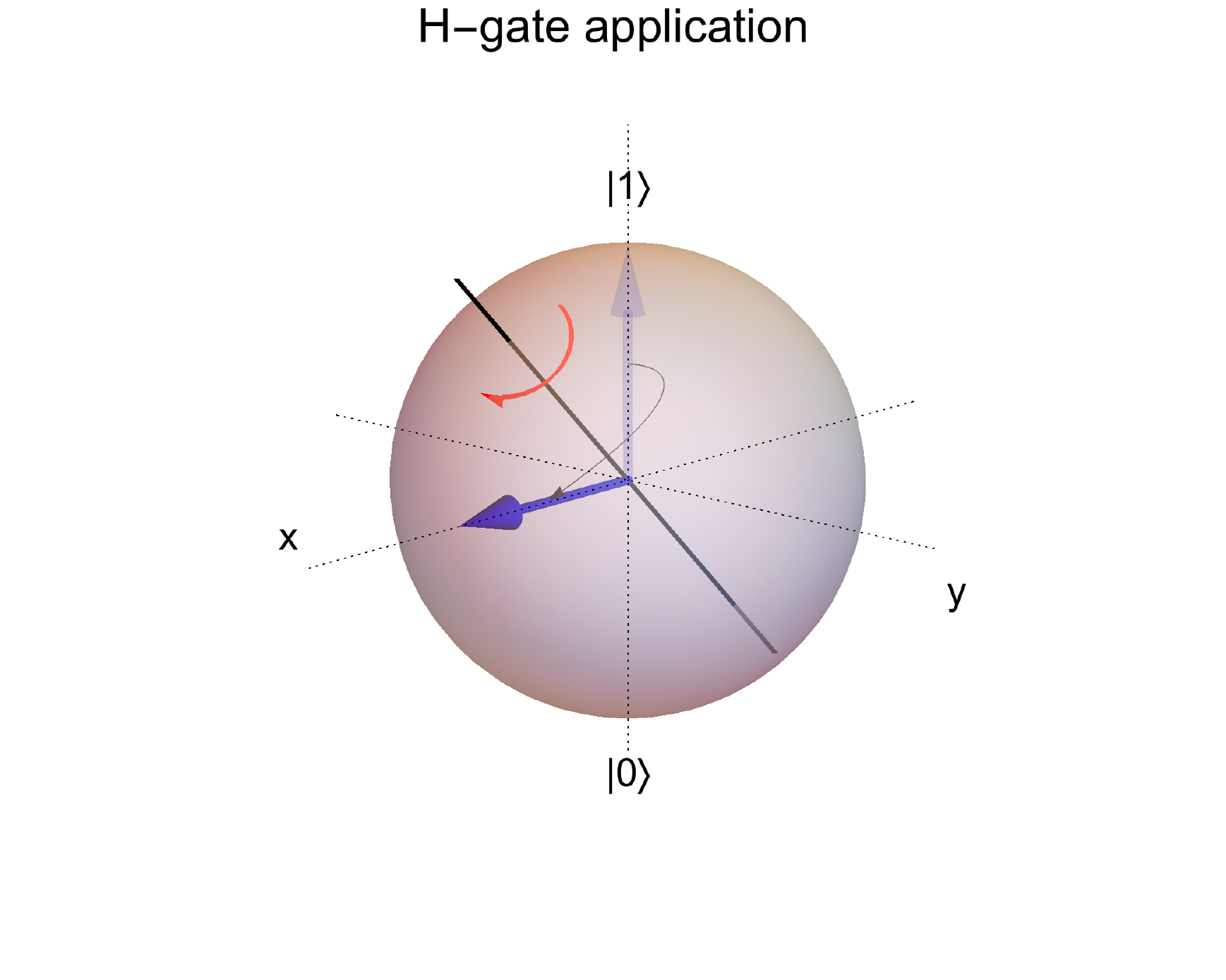}
	\hspace{-4.5 em}
        \includegraphics[height=1.9in, trim={2.5cm 0 0 0},clip]{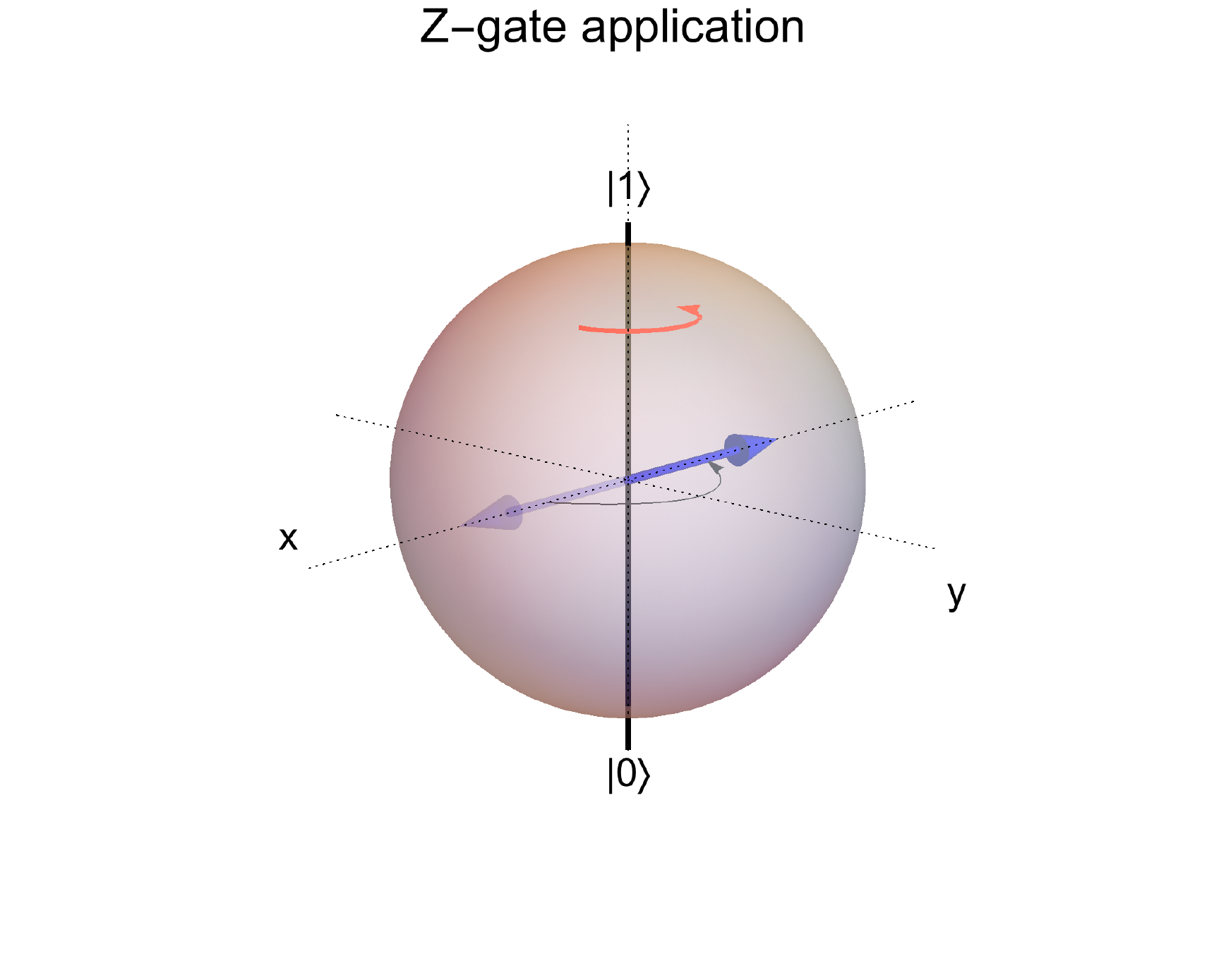}
	\hspace{-4.5 em}
	\includegraphics[height=1.9in, trim={2.5cm 0 0 0},clip]{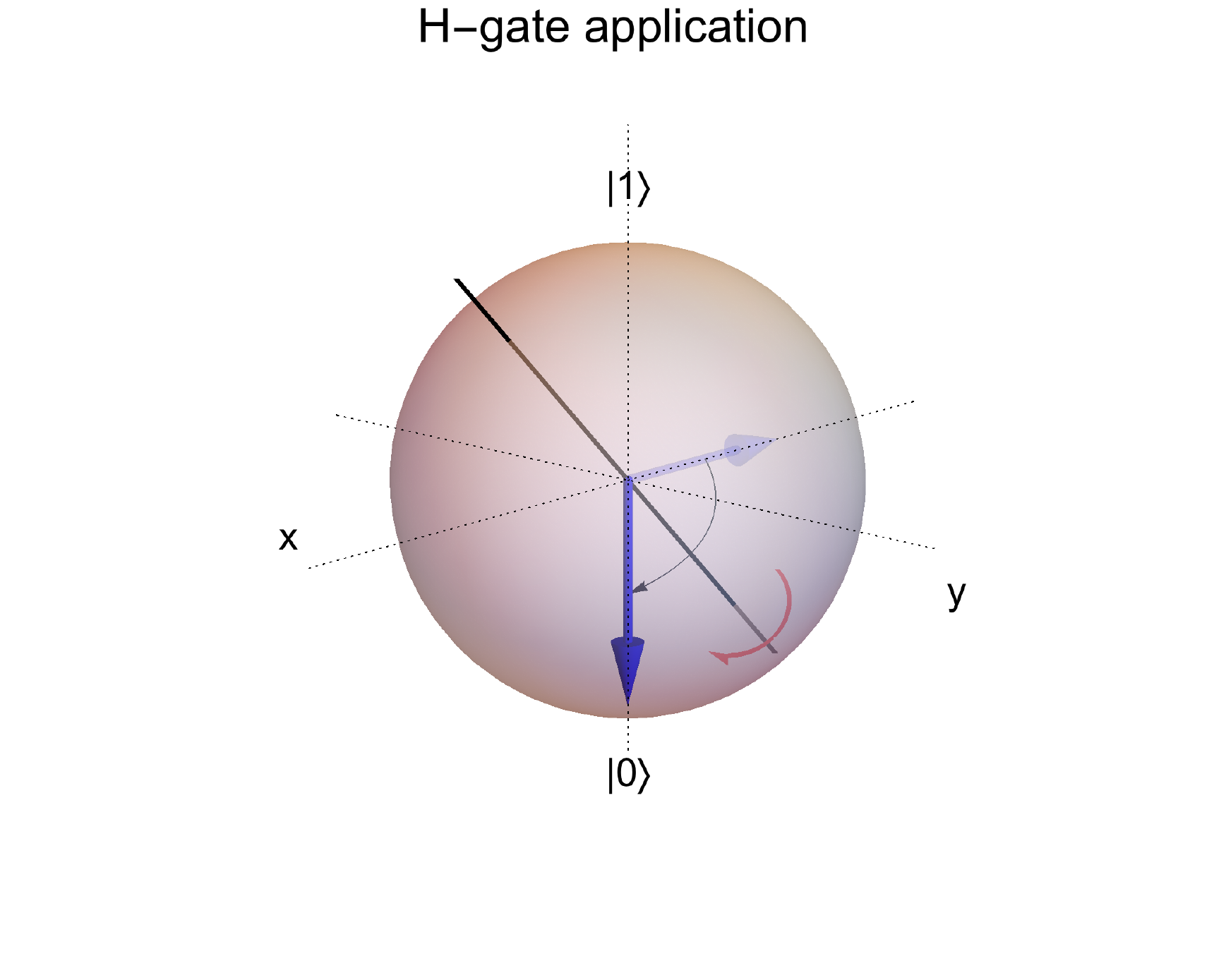}
	}
\caption{A graphical representation of the qubit state changes by applying successively three gates Hadamard, Pauli-Z and Hadamard, respectively.}
\label{fig:states}
\end{figure}

Figure~\ref{fig:states} shows that the application of these transformation to a qubit $q$ can be modelled as a rotation of the vector that represents the qubit's state around different axes. This spherical representation of the pure states of the qubits is known as a Bloch sphere. The projection of this vector on any of the three axes (i.e.~bases) corresponds to the probability of measuring that qubit as 0 or 1 on a given base.

Boolean operations can be applied on all the quantum states of the system simultaneously, meaning that several computations can happen in parallel allowing for creation of quantum algorithms that are more optimal than their classical equivalents according to the complexity theory \cite{rieffelQCintro2011}.

\subsubsection{PQCs}
\label{ssec:PQC}
A PQC \cite{mitarai2018PQC} is a circuit that consists of input qubits that are connected in a circuit of consecutive controlled parametrised gates. PQCs are typically composed of fixed gates, e.g.~controlled NOTs, adjustable gates, e.g.~qubit rotations and even at low circuit depth, some classes of PQCs are capable of generating highly non-trivial outputs \cite{benedetti2019parameterized}. 
For example consider the consecutive gates $XX^{p}$, $YY^{p}$ or $ZZ^{p}$. The first one, $XX^{p}$, is a gate equivalent to the tensor product of two $X$ gates raised to an exponent $p$, the gate parameter. Having defined $c=f\text{cos}(\pi t/2)$, $s=-if\sin(\pi t/2)$, $f=e^{i\pi t/2}$ and $w=e^{i\pi t}$, the following equations give the mathematical form of the gates:

\begin{equation}
\centering
X X^{t} 
= 
\begin{bmatrix}
c & 0 & 0 & s \\
0 & c & s & 0 \\
0 & s & c & 0 \\
s & 0 & 0 & c \\
\end{bmatrix}
\label{eq:xx}
\end{equation}

\begin{equation}
\centering
YY^{t} 
= 
\begin{bmatrix}
c & 0 & 0 & -s \\
0 & c & s & 0 \\
0 & s & c & 0 \\
-s & 0 & 0 & c \\
\end{bmatrix}
\label{eq:yy}
\end{equation}

\begin{equation}
\centering
ZZ^{t} 
= 
\begin{bmatrix}
1 & 0 & 0 & 0 \\
0 & w & 0 & 0 \\
0 & 0 & w & 0 \\
0 & 0 & 0 & 1 \\
\end{bmatrix}
\label{eq:zz}
\end{equation}

\begin{figure}
\centering
\includegraphics[width=1.02\linewidth]{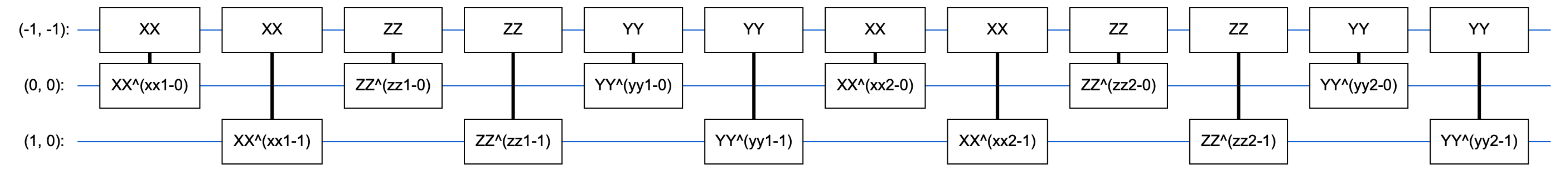}
\caption{Reduced Quantum Parametrised Circuit Digram}
\label{fig:circ_red}
\end{figure}

Figure~\ref{fig:circ_red} shows a reduced diagram of a parametrised circuit. The notation $XX^{xx1-0}$ indicates that there is a parametrised $XX$ controlled gate in the first layer (index 1) of $XX$ gates where the control qubit is a qubit with index 0 and the gate parameter is named $xx1-0$. A layer of $XX$ gates means that every input qubit has been connected to the readout qubit with a controlled parametrised $XX$ gate.

\section{Simulation Design}
\label{sect:simulations}
In this section we describe the processes of setting up of our experiments, preparing the data and designing the neural network. We represent QNNs as PQCs (Section~\ref{ssec:PQC}) in which the forward propagation is governed by quantum functions and the learned parameters are the gate parameters, such as $xx1-0$ (Figure~\ref{fig:circ_red}). These parameters are updated via the classical back propagation \cite{backPropagationrumelhart1986, lecun1989a} method which is the training part of the network. Thus, the forward propagation of these networks is fully quantum, while the backpropagation is currently classical.

Since, BiLSTM neural networks can very efficiently used for time series forecasting (Section~\ref{sect:intro}) we will be using them as a proxy for the performance of traditional deep neural networks. Note, that there are neural network models with better predictive power for time series e.g.~ the transformer-based models, Informers ~\cite{haoyietalInformer2021}, but in this work we are interested in comparing simple network structures since a robust implementation of transformer-like architectures in the form of PQCs is not currently available.

\subsection{Time series signals}
\label{ssect:simul_ts}
The time series data that we will be using for the benchmarks consist of simulated temporal signals with a varying signal-to-noise ratio (SNR). The initial objective is to benchmark the predictive power of BiLSTMs and QNNs for simulated signals representing, simplified realistic scenarios, and then proceed to more complex realistic signals.

\begin{figure}
  \includegraphics[width=0.48\linewidth]{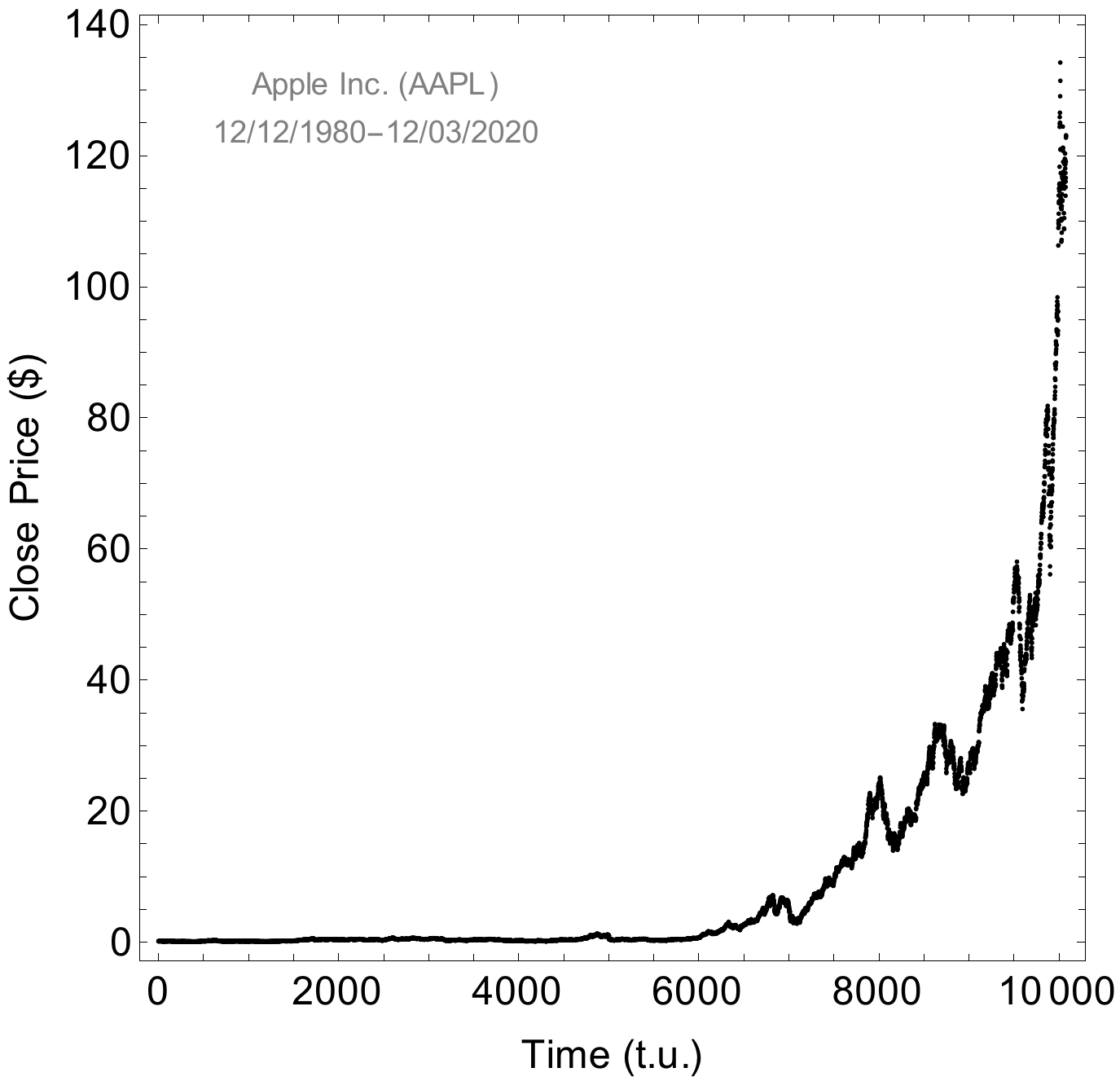}\hspace{1em}
  \includegraphics[width=0.48\linewidth]{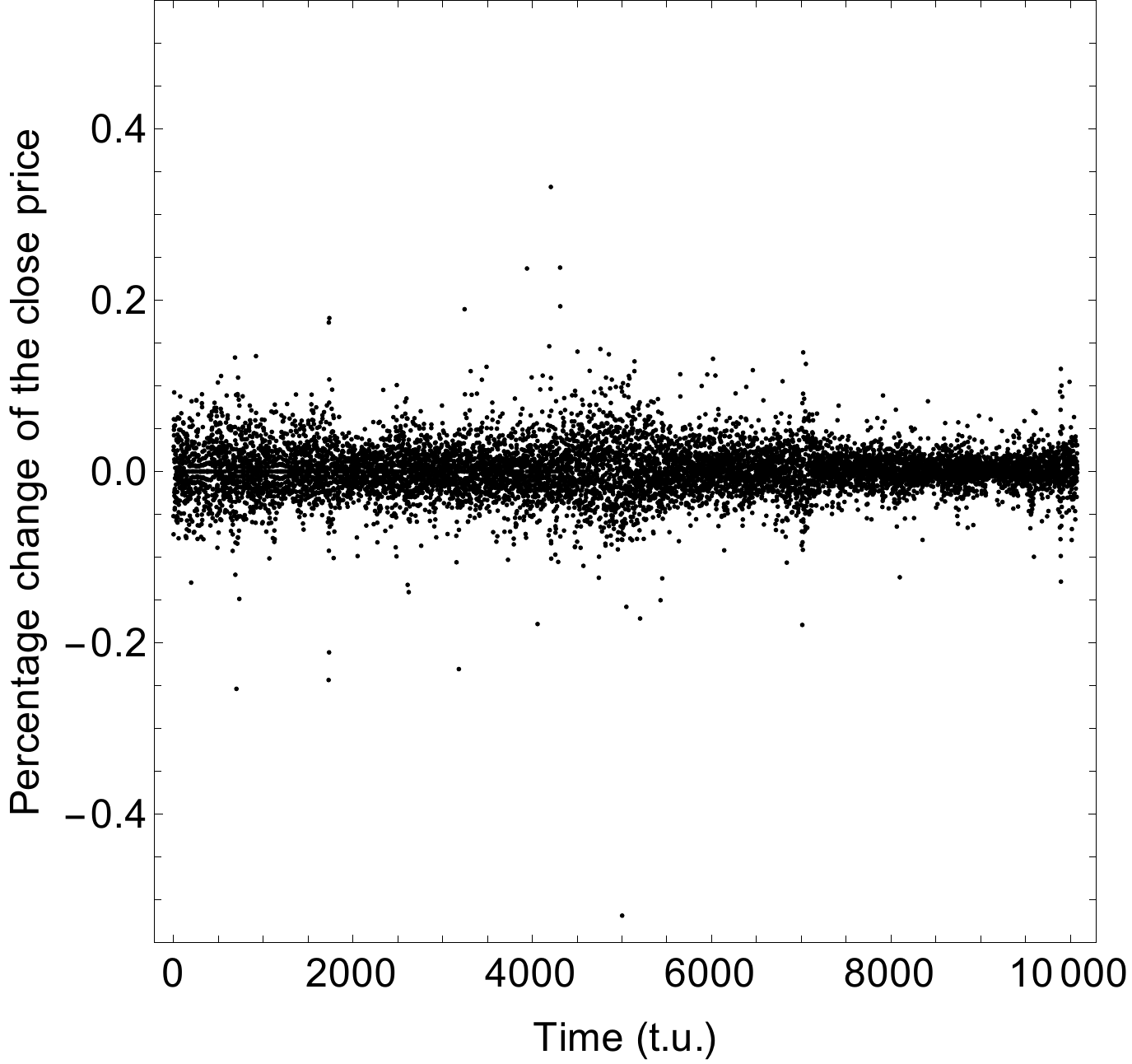}
  \caption{[Left-hand panel] The close price of the AAPL stock spanning almost 30 years.\newline[Right-hand panel] The percentage change of the close price of AAPL.}
  \label{fig:apple}
\end{figure}

\subsubsection{Simplified temporal signals}
We construct temporal signals that consist of three constituents: a fully deterministic component, a trend and a noise component.
For the deterministic part we will be using the Apple Inc. stock (AAPL) close price\footnote{The data are obtained from \href{https://finance.yahoo.com/quote/AAPL/}{Yahoo!\;Finance} consisting of $N=10080$ points and the close price is adjusted for splits.} (Figure~\ref{fig:apple}, left-hand panel) and particularly its percentage change (Figure~\ref{fig:apple}, right-hand panel). 

In order to extract from the latter a series of dominant sinusoidal components we estimate its power spectral density (or power spectrum) using the periodogram. For the purposes of this analysis we consider that all the time series are regularly sampled, there are no missing data points and they are equidistant in time with a time step $\delta t$ of 1 arbitrary time unit (t.u.). For a time series $x_i$ consisting of an even number of $N$ equidistant data points, $i=1,2,\ldots,N$ and the periodogram estimate, $P(\nu_k)$, at a given frequency $\nu_k$ of the power spectrum is the modulus-squared of the discrete Fourier transform \cite{priestley81} 

\begin{equation}
P(\nu_k)=\left|\sum_{i=0}^{N-1}x_i e^\frac{2\pi i j k}{N}\right|^2
\end{equation}
which is defined in the following $N/2+1$ positive Fourier frequencies 

\begin{equation}
\nu_k=\frac{k}{N\delta t}\;\mathrm{ with}\;k=0,1,\ldots,N/2 \;\mathrm{ and}\;\delta t=1
\end{equation}

\begin{figure}
\parbox{1\linewidth}{\includegraphics[width=0.485\linewidth]{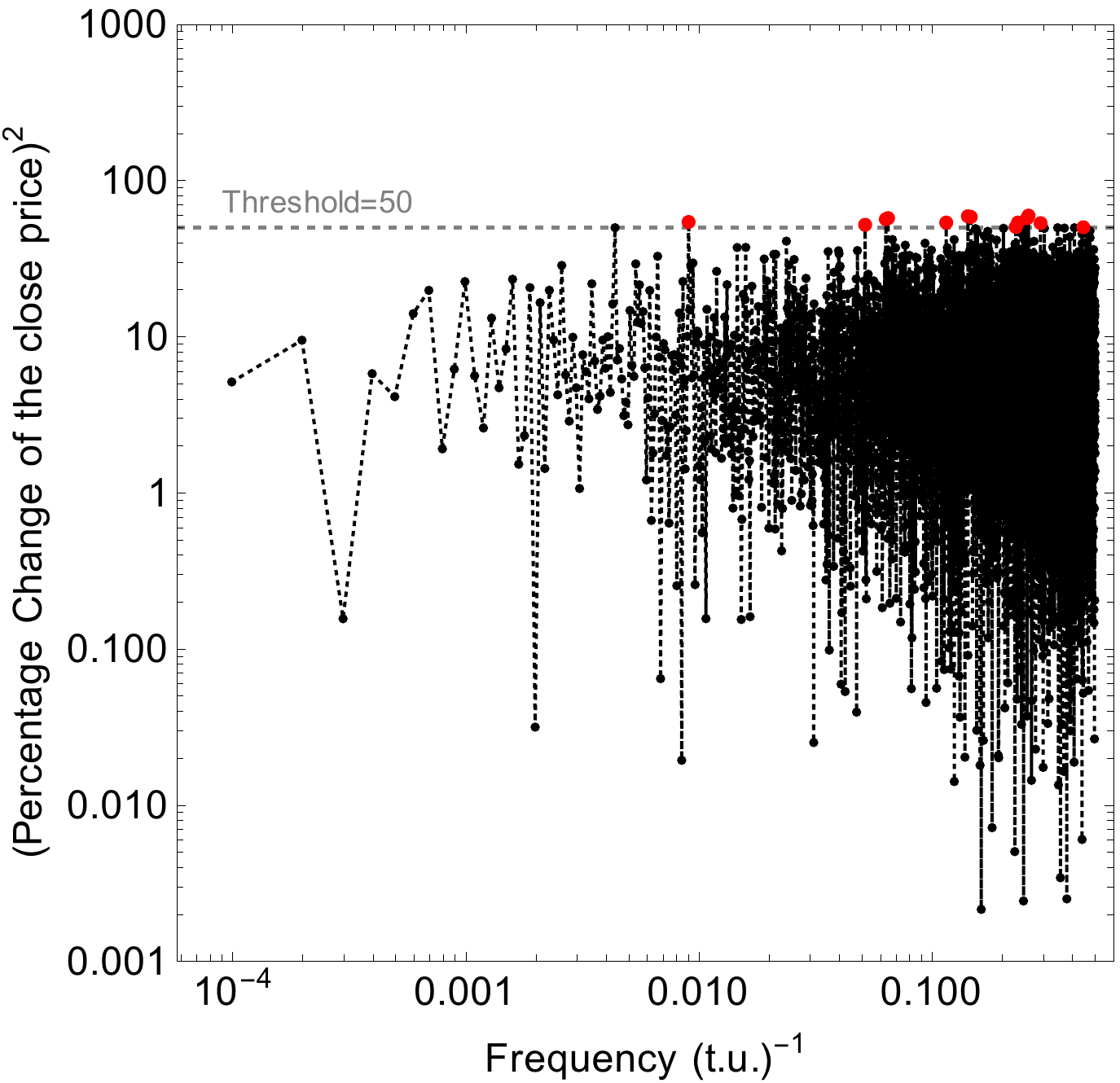}
\parbox{1\linewidth}{\vspace{-17.5em}\includegraphics[width=0.52\linewidth]{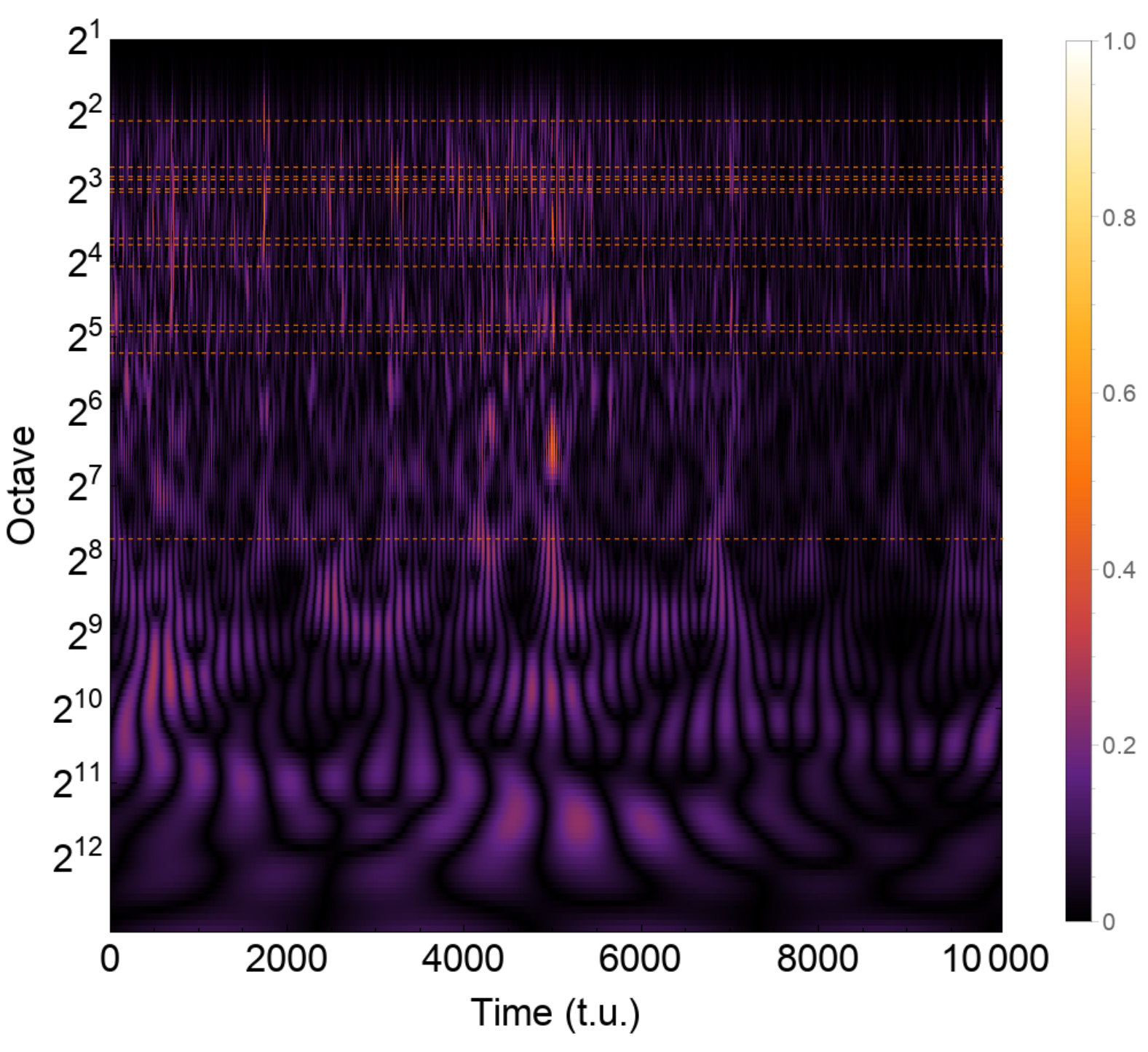}}
}
\caption{Temporal analysis of the percentage change for the closing price of AAPL. [Left-hand side] The periodogram estimates of the time series. The 13 red points indicate the sinusoidal components whose squared amplitude exceeds 50.\newline[Right-hand side] The continuous wavelet transform of the time series using as a wavelet the derivative of Gaussian wavelet with a derivative order of 12. The wavelet coefficients have been estimated for 12 octaves with 12 voices at each octave and the red dotted lines indicate the position of the 13 sinusoids whose squared amplitude exceeds 50 (i.e.~the red points of the left-hand panel of this figure.}
\label{fig:psdWaveletApple}
\end{figure}

In the left-hand side of Figure~\ref{fig:psdWaveletApple} we show the periodogram estimates of the percentage change of the close price, indicating with the 13 red points the sinusoidal components whose sqared amplitude exceeds 50, which is an arbitrary set threshold that we have selected for this analysis. By bringing up or down this threshold one can include less or more sinusoidal components, respectively.

The values of these 13 sinusoidal components are given in Table~\ref{table:determPart} and their sum yields the deterministic part of our signal.

\begin{table}
\centering
\begin{tabular}{c c}
\hline
Ampitude ID & Frequency \\ \hline\hline
7.37 & 0.009 \\
7.22 & 0.052 \\
7.52 & 0.063 \\
7.58 & 0.065 \\
7.33 & 0.115 \\
7.68 & 0.143 \\
7.65 & 0.146 \\
7.10 & 0.229 \\ 
7.34 & 0.233 \\
7.58 & 0.256 \\
7.72 & 0.259 \\
7.30 & 0.292 \\
7.09 & 0.445 \\ \hline
\end{tabular}
\caption{The 13 sinusoidal components derived from the the percentage change for the closing price of AAPL having a squared amplitude greater than 50.}
\label{table:determPart}
\end{table}

In order to check (i) if the time series of the percentage change of the close price consists of intermittent events, and if this is the case, (ii) if any of these sinusoidal components coincides with them we perform a wavelet analysis e.g.~\cite{rioul1991wavelets}. We are using the derivative of Gaussian wavelet \cite{tu2005analysis} of derivative order $k=12$ which has the following mother wavelet transforming function
\begin{eqnarray}\nonumber
\psi_k(t)&=&\frac{(-1)^{k+1}}{\sqrt{\Gamma\left(k+\frac{1}{2}\right)}}\frac{d^k}{dx^k}e^{-\frac{t^2}{2}}\stackrel{k=12}{\Longrightarrow}\\ 
\psi_{12}(t)&=&-\frac{64 e^{-\frac{t^2}{2}} \left(t^{12}-66 t^{10}+1485 t^8-13860
   t^6+51975 t^4-62370 t^2+10395\right)}{315 \sqrt{3187041} \sqrt[4]{\pi}}
\end{eqnarray}
in which $\Gamma\left(k+\frac{1}{2}\right)$ corresponds to the (complete) gamma function.

The continuous wavelet transform is given by
\begin{equation}
w(u,s)=\frac{1}{\sqrt{s}}\sum_{i=1}^N x_i \psi_{12}^*\left(\frac{\delta t(i-u)}{s}\right)
\end{equation}
in which the asterisk represents the operation of complex conjugation, $\delta t=1$, and $s$ is the scaling parameter given by $\alpha 2^{oct-1}2^{voc/nvoc}$ where $\alpha$ is the smallest wavelet scale, $oct$ and $voc$ is the octave and voice, respectively, and $noct$ and $nvoc$ the total number of octaves and voices, per octave, respectively. 

In the right-hand side of Figure~\ref{fig:psdWaveletApple} we show the continuous wavelet transform of the percentage change of the close price of AAPL using $noct=12$ octaves with $nvoc=12$ voices per octave. The red dotted lines correspond to the 13 frequency components, depicted by the periodogram analysis, whose squared amplitude is more than 50 (Figure~\ref{fig:psdWaveletApple}, left-hand side). As we can see, for these sinusoids during the entire time span, their amplitude does not fluctuate dramatically (the fluctuations are of the order of 23 per cent) meaning that the majority of the time they do sample the average behaviour of the underlying process and thus they are representative for the large amplitude fluctuations of the signal.

Having composed the deterministic component of the signal, consisting of the 13 sinusoidal constituents, we proceed to the construction of the second and the third signal components which are the trend and the noise. These two components (i.e.~the distortion components) are the ones that we will be changing in order to make the signal gradually more complex. In total we produce 15 different versions of these combined components and their values are given in Table~\ref{table:signals} in which F, L and Q stand for flat, linear and quadratic trend, respectively. The noise component is generated by a Uniform distribution, $U(0,1)$. Finally, the signals are scaled by multiplying the noise signal with noise coefficient calculated as absolute difference between the maximum and minimum values of the sinusoid signal with and without trend. For each signal, we simulate 10,000 time steps measured in arbitrary time units (t.u.) with regular sampling without any missing data points.

\begin{table}
\centering
\begin{tabular}{c c c c c }
\hline
Signal ID & \vtop{\hbox{\strut Linear gradient}\hbox{\strut \hspace{1.5em}($\times 10^{-2}$)}} & \vtop{\hbox{\strut Quadratic gradient }\hbox{\strut \hspace{2.5em}($\times 10^{-5}$)}} & Noise coefficient & SNR                   \\ \hline\hline
F0        & 0               & 0                  & 0.0                                           & 0.0005              \\
F2        & 0               & 0                  & 0.2                                           & 0.6603              \\
F5        & 0               & 0                  & 0.5                                           & 1.2445              \\
F8        & 0               & 0                  & 0.8                                           & 1.4716              \\ 
F10       & 0               & 0                  & 1.0                                           & 1.5472              \\
L0        & 5               & 0                  & 0.0                                           & 1.1793              \\ 
L2        & 5               & 0                  & 0.2                                           & 2.0653              \\
L5        & 5               & 0                  & 0.5                                           & 2.3653              \\ 
L8        & 5               & 0                  & 0.8                                           & 2.4303              \\ 
L10       & 5               & 0                  & 1.0                                           & 2.4251              \\ 
Q0        & 0               & 5          	 & 0.0                                           & 1.1179              \\ 
Q2        & 0               & 5          	 & 0.2                                           & 1.4323              \\ 
Q5        & 0               & 5          	 & 0.5                                           & 1.7597              \\ 
Q8        & 0               & 5           	 & 0.8                                           & 1.9498              \\ 
Q10       & 0               & 5           	 & 1.0                                           & 2.0125              \\ 
\end{tabular}
\caption{The 15 distortion components which are added individually to the deterministic signal consisting of the combined sinusoids given in Table~\ref{table:determPart}.}
\label{table:signals}
\end{table}

\subsubsection{Realistic temporal signals}
\begin{figure}
  \includegraphics[width=0.50\linewidth]{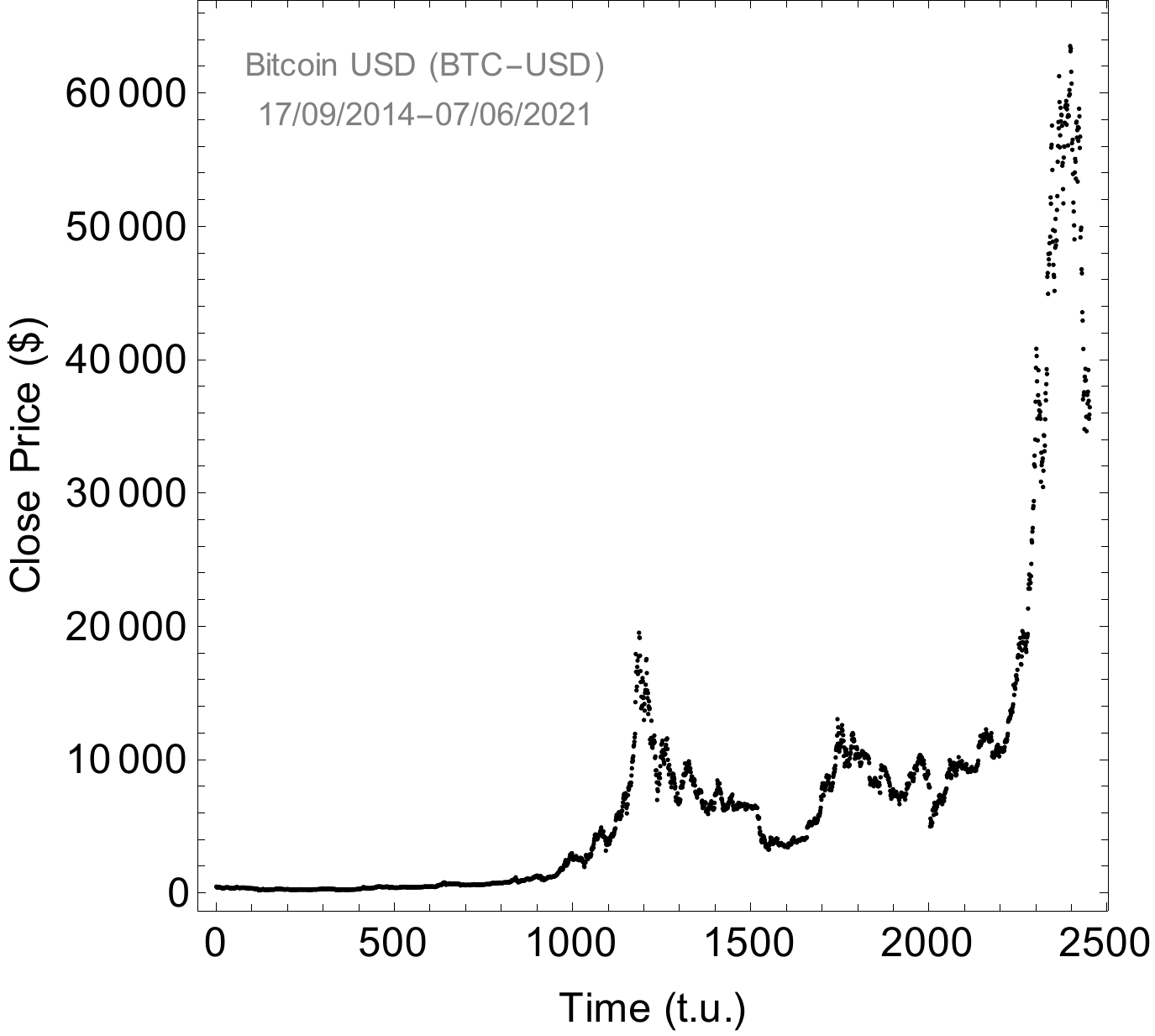}\hspace{1em}
  \includegraphics[width=0.48\linewidth]{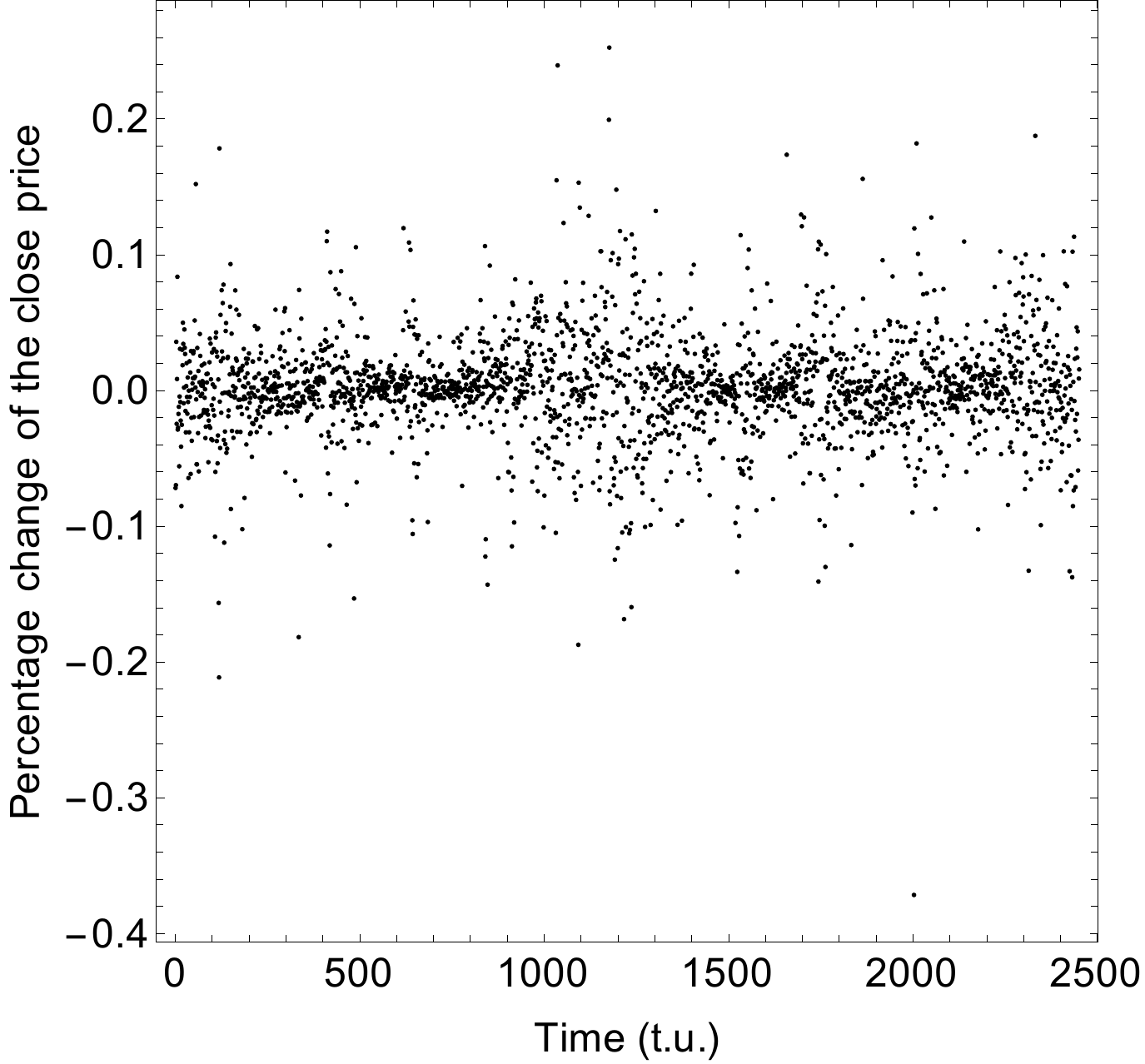}
  \caption{[Left-hand panel] The close price of the BTC-USD stock spanning almost 7 years.\newline[Right-hand panel] The percentage change of the close price of BTC-USD.}
  \label{fig:bitcoin}
\end{figure}

\begin{figure}
  \includegraphics[width=0.49\linewidth]{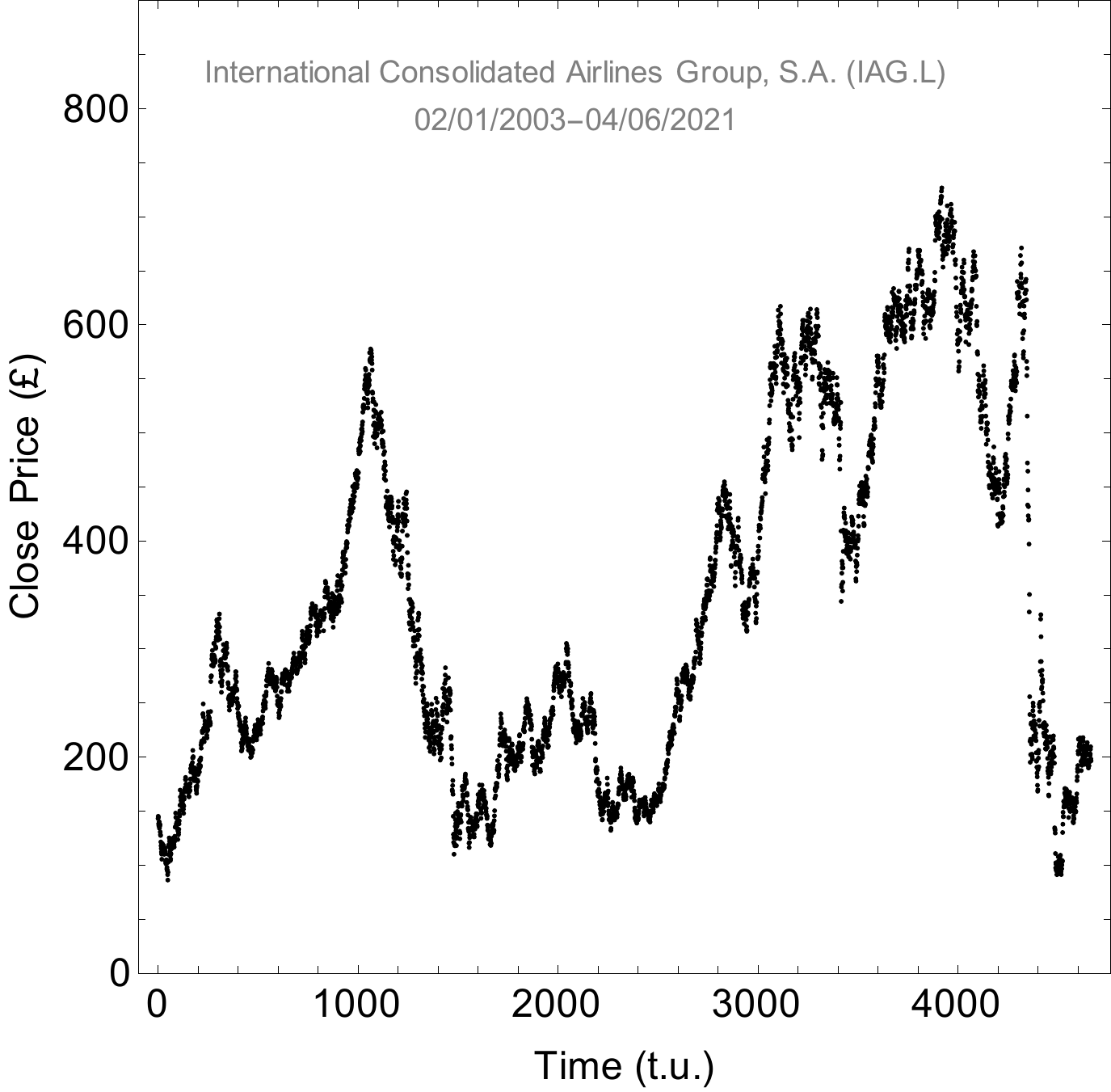}\hspace{1em}
  \includegraphics[width=0.50\linewidth]{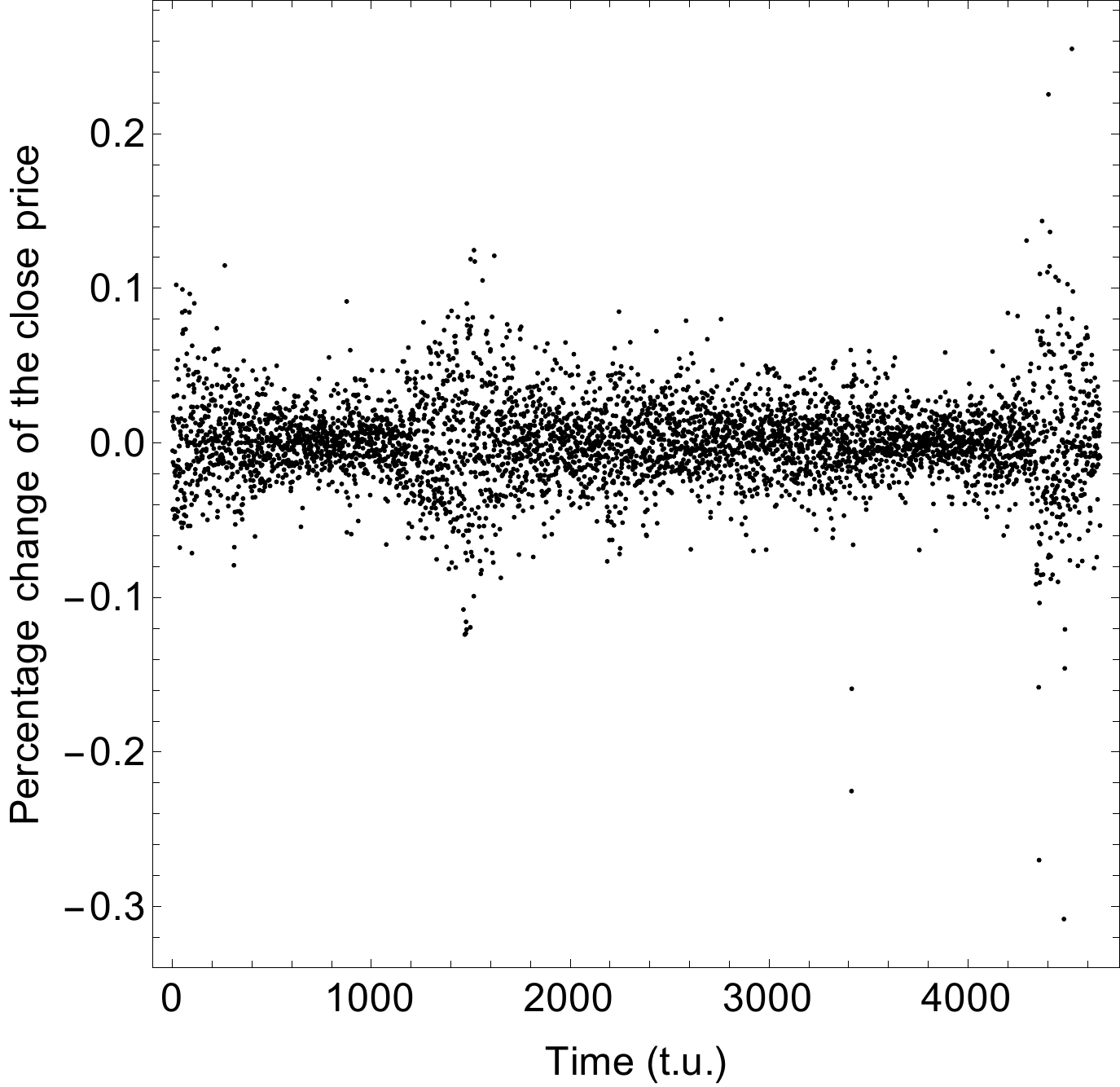}
  \caption{[Left-hand panel] The close price of the IAG.L stock spanning almost 18.5 years.\newline[Right-hand panel] The percentage change of the close price of IAG.L.}
  \label{fig:iagl}
\end{figure}

After testing the predictive power both the QNNs and BiLSTM neural network with simple temporal signals, we proceed with more complex and realistic signals. Initially, we consider the entire signal of the percentage change of the AAPL stock as it is shown in the right-hand panel of Figure~\ref{fig:apple}. Then we consider the percentage change of Bitcoin USD (BTC-USD) (Figure~\ref{fig:bitcoin}, right-hand panel), derived from the 2452 points of its closing prices (Figure~\ref{fig:bitcoin}, left-hand panel) and the percentage change of the International Consolidated Airlines Group, S.A. (IAG.L) (Figure~\ref{fig:iagl}, right-hand panel), derived from the 4666 points, of its closing prices (Figure~\ref{fig:iagl}, left-hand panel).

\subsection{Data pre-processing}
\label{ssect:datPreProc}
Initially, the data must be partitioned in order to create the model input and output. Thus, for a model input of size $m$, the signal would be sampled in segments of length $m+1$. Each segment contains a sequence of $m$ values as input; the first value corresponds to the value of the signal at time 0 (t.u.), for that segment, and the last value corresponds to the value at time $m$ (t.u.) which is the target output value of the model. 

The first three quarters of the signal are used for training and validation and the last quarter is used for testing. Each of these two segments is split into groups of 17 consecutive points which are then split into 16 points for model input and 1 point for model output i.e.~the forecasting horizon is 1 (t.u.).

The \texttt{MinMaxscaler} with parameters 0.2 as a lower bound and 0.8 as an upper bound of the scaling range is used for the training data. The scaler is fitted to the training data samples and then it is used to transform both the training and testing data samples. This allows for test values outside the range of the train data in the margin of 0.2 to be scaled between 0 and 1. Here, it is important to account for signals with significant trends that will distort the resolution of the scale. The data are scaled in the same way for both the PQC and the BiLSTM inputs, but the data for the QNN need also to be encoded to a quantum array.

\subsection{Quantum encoding and PQC design}
In order the PQCs to be used as a time series forecast tool, the signal needs to be encoded into qubits. Since, the classical $m$ data points in each sample are scaled between 0 and 1, they can be encoded in a qubit array of the same size. In order to preserve the resolution of the samples and give physical meaning to the values in one sample relative to each other the input of 16 values is centred around 0.5 and each value is then encoded to a 0 qubit with a gate $X^a$ where $a$ is the classical value between the 0 to 1 range. All the values in the sample that are larger than the median have a higher probability to be measured as 1 rather than 0 and vice versa. The predicted output is calculated as the expected value of the readout qubit which is always between the 0 to 1 range. Finally, for the testing, the output is inversely transformed with the original \texttt{MinMaxscaler} (Section~\ref{ssect:datPreProc}) yielding the final predicted values.

The parametrised quantum circuit that is used as a neural network to exhibit the potential for forecasting time series consist of 6 layers of gates in which each layer has a controlled parametrised gate connecting each input qubit to the readout qubit. Considering that there are 16 input qubits and 6 layers of the network, the total number of parametrised gates and total number of parameters is 96. The 6 layers of gates are the following: $XX^{xx1}$, $ZZ^{zz1}$, $YY^{yy1}$,  $XX^{xx2}$, $ZZ^{zz2}$, $YY^{yy2}$. Note, that this combination of gates was selected based on model accuracy after experimenting with different combinations of layers.

\subsection{BiLSTM neural network}
\begin{figure}
  \centering
  \includegraphics[width=1.015\linewidth,trim={0 0 92 0},clip]{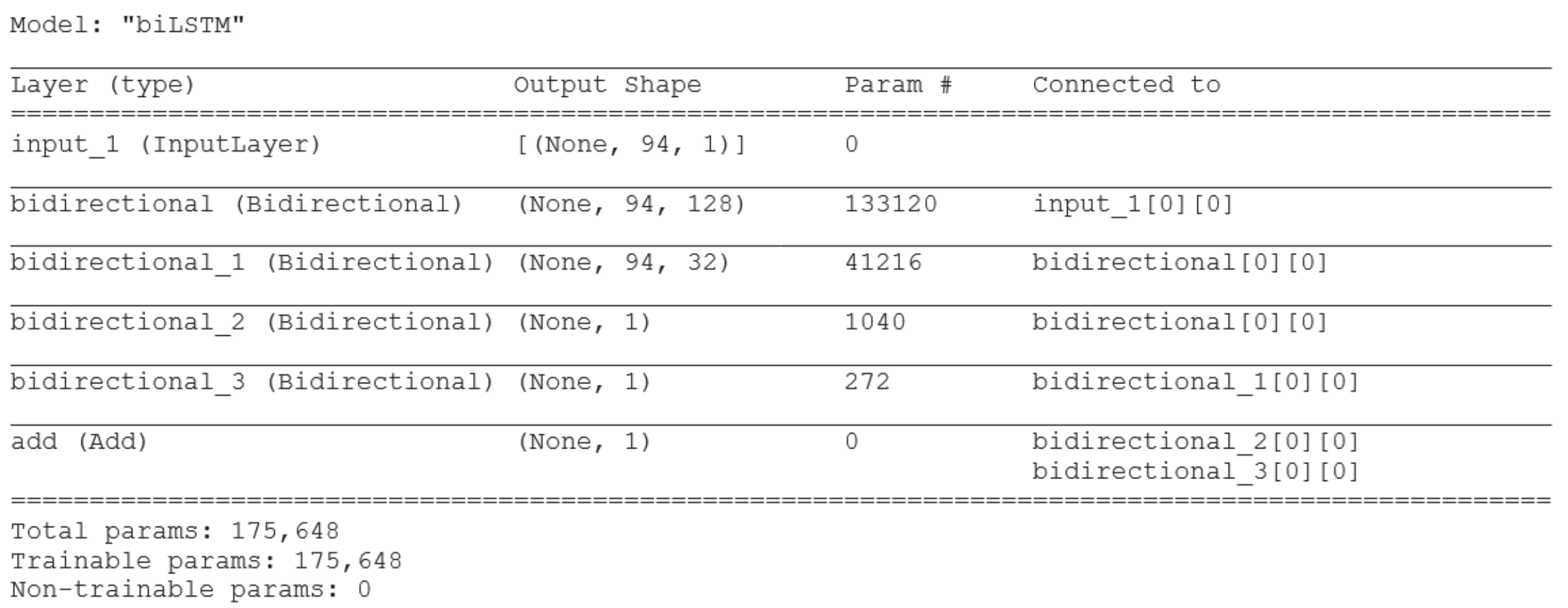}
  \caption{BiLSTM neural network model summary (report produced by TensorFlow).}
  \label{fig:lstm}
\end{figure}

The BiLSTM neural network that we use in this experiment consists of 4 bidirectional layers resulting in 175,648 trainable parameters (Figure~\ref{fig:lstm}). The first layer \textit{bd\_seq} after the input is made of 128 units and \textit{tanh} activation function. The output of this layer is fed as an input to 2 different layers \textit{bd\_sin}, the one having 32 units and \textit{tanh} activation function and \textit{bd\_1} that the other one has 1 unit and \textit{linear} activation function. The output of the \textit{bd\_sin} layer is then fed to a layer \textit{bd\_2} with one unit and \textit{tanh} activation function. The output is calculated as addition of the \textit{bd\_1} and \textit{bd\_2} 
layers. The BiLSTM neural network model is trained using the \textit{Adam} optimiser \cite{kingma2014adam} with mean squared error used as a loss function.

\section{Comparison: QNNs vs. BiLSTM neural network}
\label{sect:comparison}
\begin{figure}
\hspace{-1em}
\parbox{1.1\linewidth}{\includegraphics[width=0.48\linewidth]{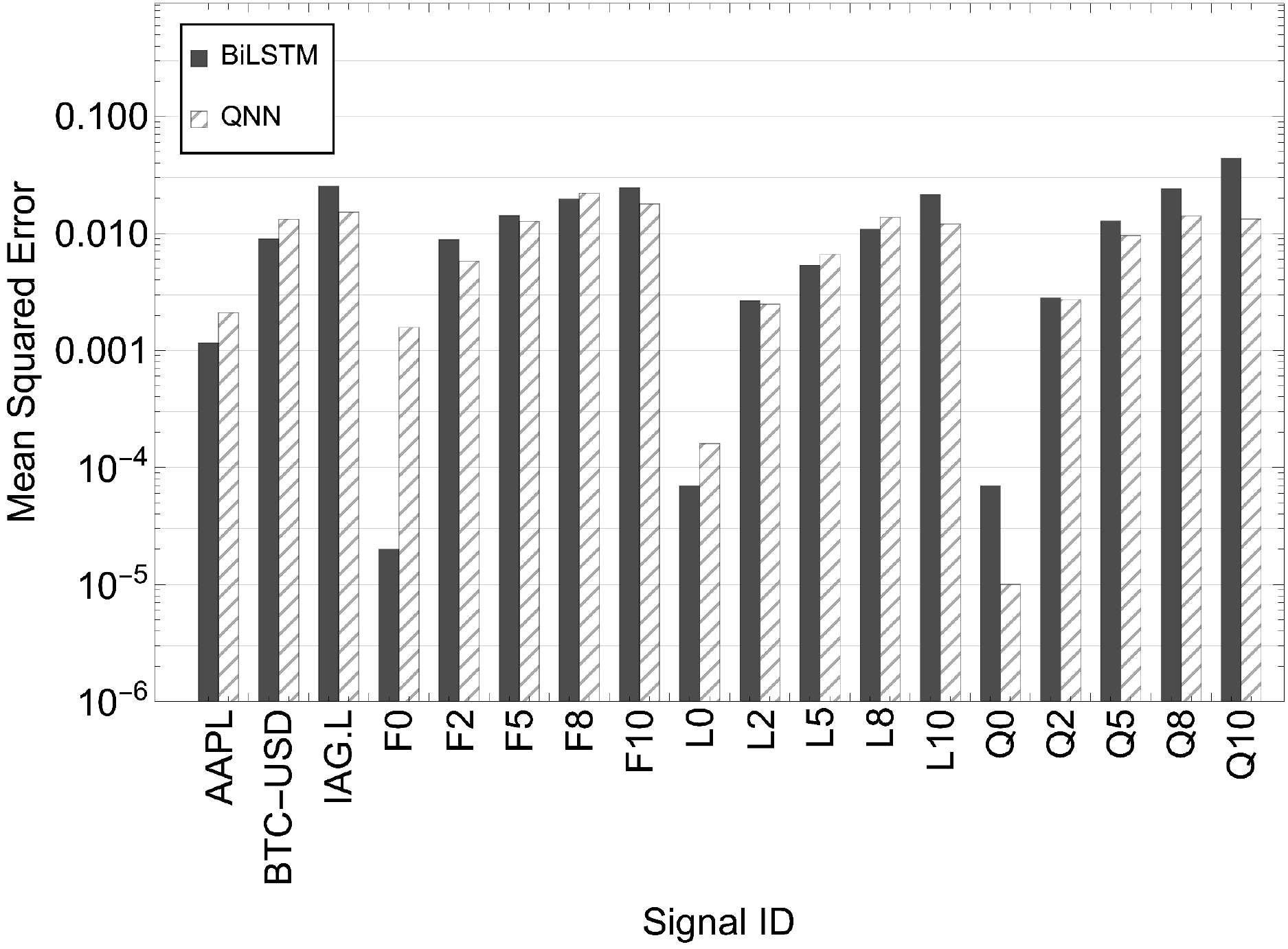}
\includegraphics[width=0.48\linewidth]{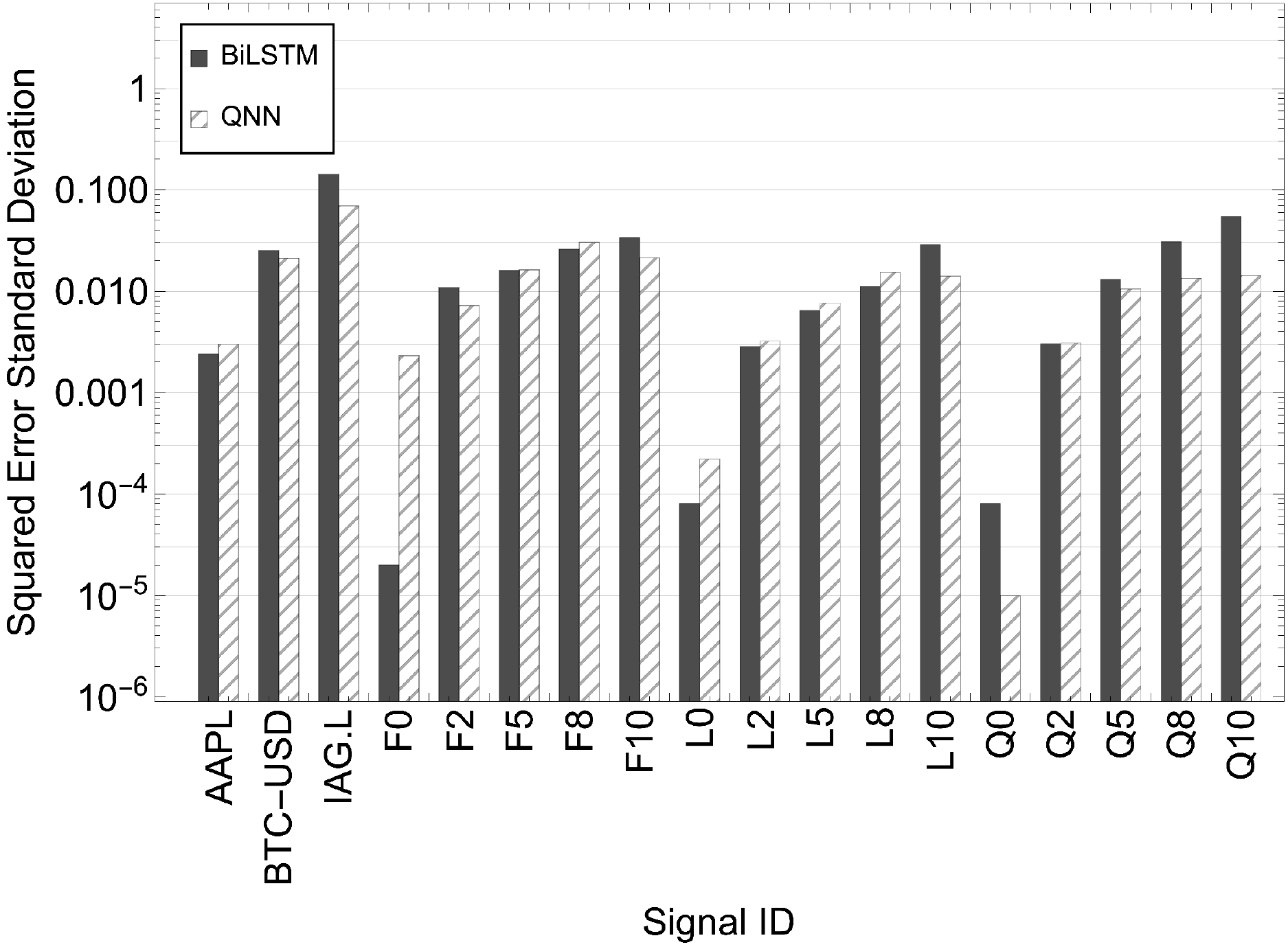}}
\hspace{-1em}
\parbox{1.1\linewidth}{\includegraphics[width=0.48\linewidth]{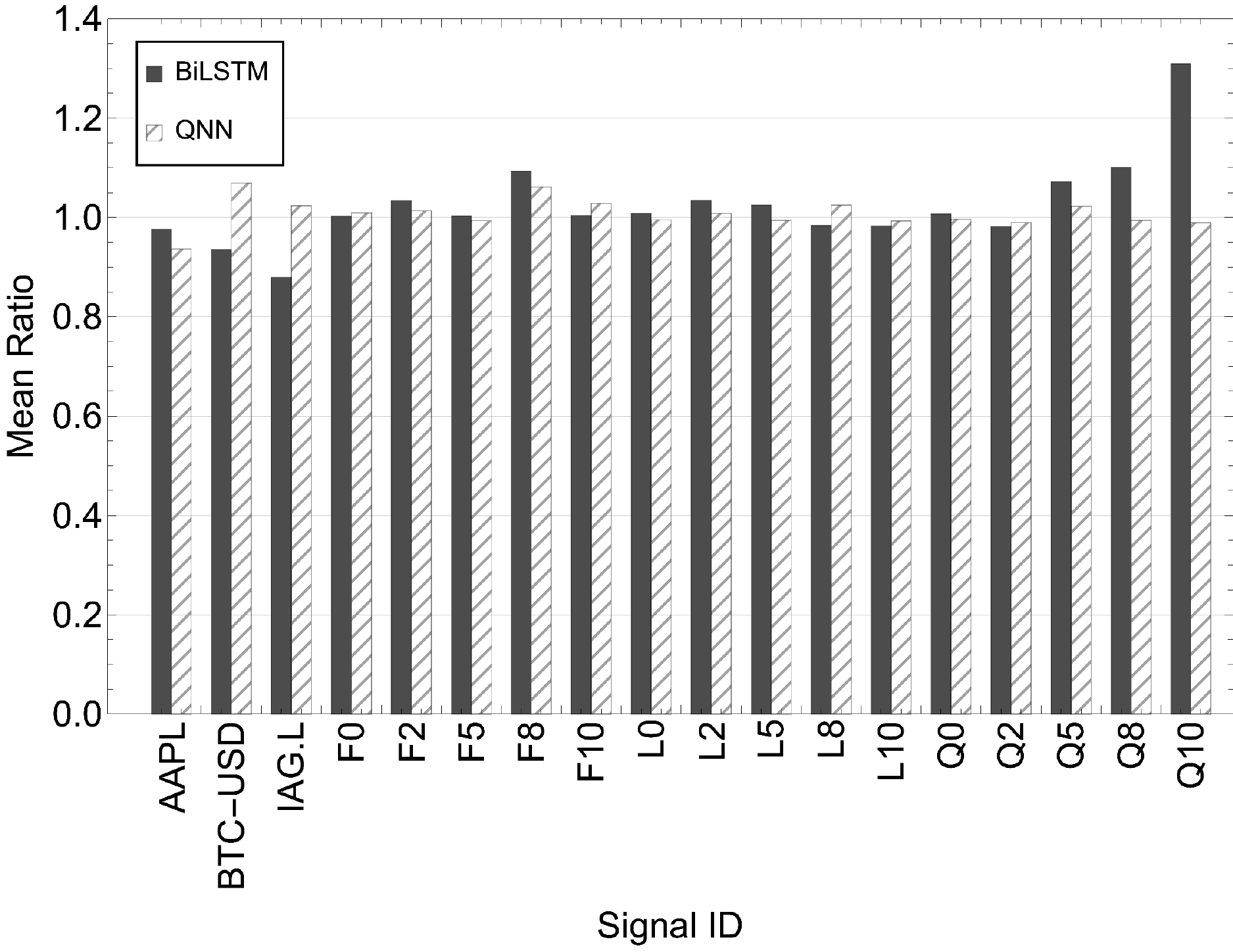}
\includegraphics[width=0.48\linewidth]{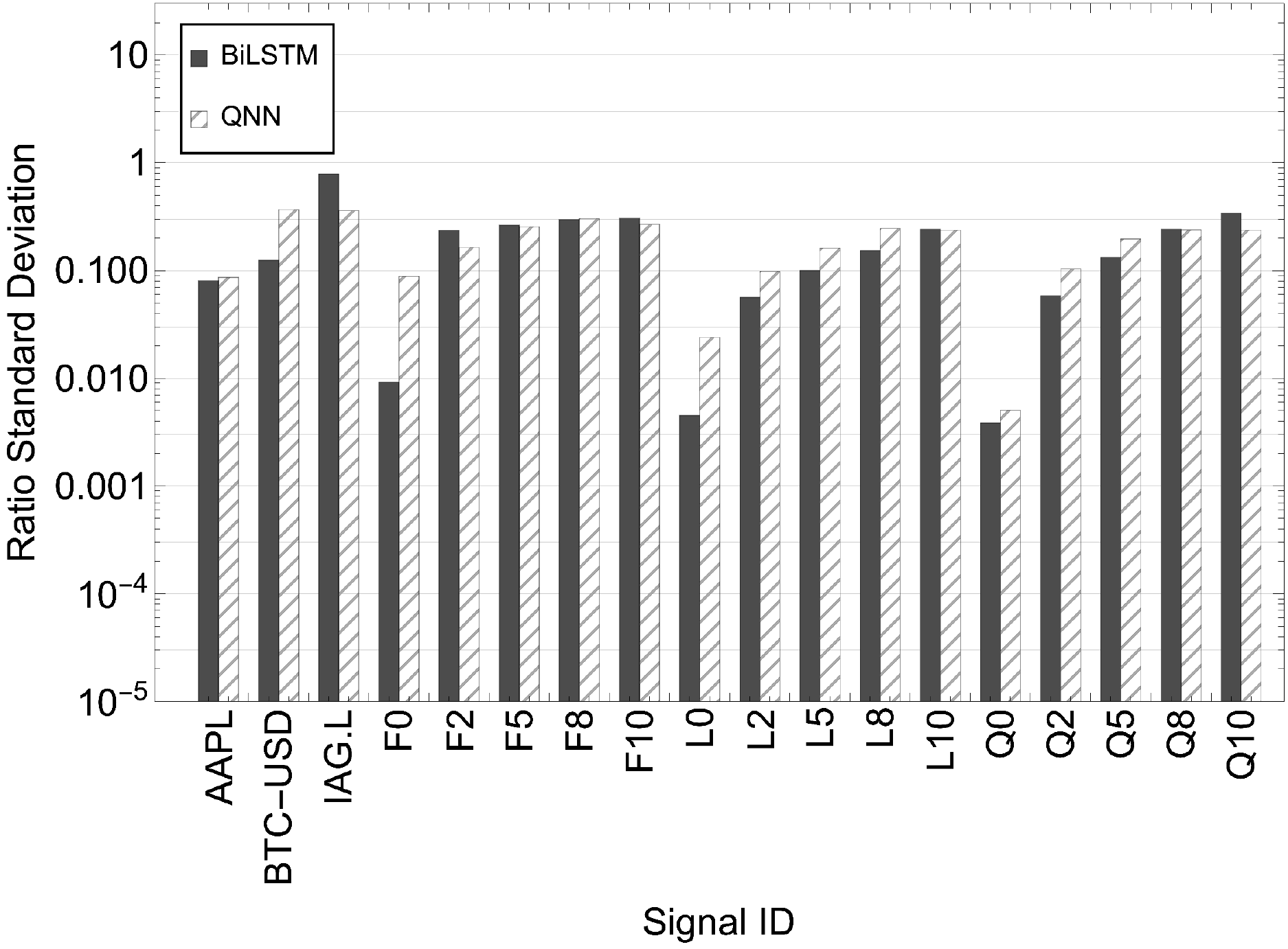}}
\caption{The comparison results of the metrics between QNN and BiLSMT neural network as derived from the simulated test data sets.}
\label{fig:results}
\end{figure}

Both the QNNs and the BiLSTM neural networks are trained for 300 epochs for all the different types of simulated time series signals (Section~\ref{ssect:simul_ts}) and their forecasting power is compared in this section. In Table~\ref{table:results} and Figure~\ref{fig:results} we show the results using five metrics for the test data set: the mean square error (MSE), the square error standard deviation (SESD), the mean ratio (MR) and the standard deviation of the ratio (SDR). 

These results show that the PQCs, as a QNN equivalent, can perform similar to the BiLSTM neural network in all cases and slightly outperform it on signals with higher noise coefficient. Also, the PQCs have less tendency to overfit the noise as it consist only of 96 trainable parameters compared to the BiLSTM neural network that has 175,648 trainable parameters. This behaviour can be understood in the context of the sinusoidal nature of the signals and the types of functions that govern the two models. The signals are functions with trigonometric patterns and the functions that govern the quantum circuit are also trigonometric, which allows the circuit to fit (with fewer parameters) the signal. On the other hand, the classical neural network is governed by polynomial functions (plus sigmoid and tangent as activation functions), implying many polynomial parameters would be required to approximate and finally fit a trigonometric signal.

\begin{table}[h!]
 \centering
\begin{tabular}{l l l l l l}
\hline
Signal ID             & NN & MSE & SESD & MR & SDR \\ \hline\hline
\multirow{2}{*}{F0}   & BiLSTM  & \textbf{0.00002}   &  \textbf{0.00002}   &  \textbf{1.00289}  &  \textbf{0.00922}     \\ 
                      & QNN  &  0.00157   &  0.00230    &  1.00909  & 0.08807   \\ \hline
\multirow{2}{*}{F2}   & BiLSTM  & 0.00888    & 0.01089    & 1.03396   &  0.23622  \\ 
                      & QNN  &  \textbf{0.00578}   &  \textbf{0.00724}  &  \textbf{1.01314}  & \textbf{0.16389} \\ \hline
\multirow{2}{*}{F5}   & BiLSTM  &  0.01421   &  \textbf{0.01600}   &  1.00345  &   0.26243 \\ 
                      & QNN  &  \textbf{0.01267}   &  0.01625   &  \textbf{0.99319}  &  \textbf{0.25542}  \\ \hline
\multirow{2}{*}{F8}   & BiLSTM  &  \textbf{0.01980}   &  \textbf{0.02590}   &  1.09280  & \textbf{0.29637}   \\ 
                      & QNN  &  0.02197   & 0.03032    &  \textbf{1.06121}  & 0.30332   \\ \hline
\multirow{2}{*}{F10}  & BiLSTM &  0.02461   & 0.03384    & \textbf{1.00418}   & 0.30623   \\ 
                      & QNN  &  \textbf{0.01786}   & \textbf{0.02133}    &  1.02801  & \textbf{0.26719}   \\ \hline
\multirow{2}{*}{L0}   & BiLSTM  &  \textbf{0.00007}   &  \textbf{0.00008}   &  1.00841  &  \textbf{0.00453}  \\ 
                      & QNN  &   0.00016  &  0.00022   & \textbf{0.99553}   & 0.02403   \\ \hline
\multirow{2}{*}{L2}   & BiLSTM &  0.00265   &  \textbf{0.00285}   & 1.03427   &  \textbf{0.05614}  \\ 
                      & QNN  & \textbf{0.00248}    &  0.00321   & \textbf{1.00801}   &  0.09831  \\ \hline
\multirow{2}{*}{L5}   & BiLSTM  &   \textbf{0.00531}  &  \textbf{ 0.00647}  &  1.02451  & \textbf{0.10029}  \\ 
                      & QNN  & 0.00658    &  0.00767  & \textbf{0.99409}   & 0.16051   \\ \hline
\multirow{2}{*}{L8}   & BiLSTM &  \textbf{0.01091}   &  \textbf{0.01107}   & \textbf{0.98402}   &   \textbf{0.15356} \\ 
                      & QNN  & 0.01366    &  0.01541   &  1.02479  &  0.24523   \\ \hline
\multirow{2}{*}{L10}  & BiLSTM &  0.02150   &  0.02864   &  \textbf{0.98316}  &  0.24226  \\ 
                      & QNN  & \textbf{0.01205}    & \textbf{0.01404}    &  0.99251  &  \textbf{0.23725}  \\ \hline
\multirow{2}{*}{Q0}   & BiLSTM  &  0.00007   &  0.00008   &  1.00712  &  \textbf{0.00385}  \\ 
                      & QNN  &  \textbf{0.00001}   &  \textbf{0.00001}   & \textbf{0.99675}   &  0.00506  \\ \hline
\multirow{2}{*}{Q2}   & BiLSTM  & 0.00282    & \textbf{0.00305}    & \textbf{0.98216}   & \textbf{0.05855}   \\ 
                      & QNN  &   \textbf{0.00273}  & 0.00309    &  0.98988  &  0.10283  \\ \hline
\multirow{2}{*}{Q5}   & BiLSTM &  0.01278   &  0.01310   & 1.07170   &  \textbf{0.13308}  \\ 
                      & QNN  &  \textbf{0.00962}   & \textbf{0.01056}    &   \textbf{1.02279}  &  0.19553  \\ \hline
\multirow{2}{*}{Q8}   & BiLSTM  &  0.02414   &  0.03065   &  1.10128  &  0.24244  \\ 
                      & QNN  &  \textbf{0.01415}   &  \textbf{0.01345}   & \textbf{0.99407}   &  \textbf{0.23837}  \\ \hline
\multirow{2}{*}{Q10}  & BiLSTM  & 0.04381   &  0.05485   & 1.30926  & 0.33938 \\ 
                      & QNN  &   \textbf{0.01333}  &  \textbf{0.01427}   &  \textbf{0.98959}  & \textbf{0.23724}   \\ \specialrule{1.5pt}{1pt}{1pt}
\multirow{2}{*}{AAPL} & BiLSTM  &  \textbf{0.00116}   &  \textbf{0.00242}  & 0.97601 &  \textbf{0.08056} \\ 
                      & QNN  &  0.00211   &  0.00301   &  \textbf{0.93657}  & 0.08659  \\ \hline
\multirow{2}{*}{BTC-USD}   & BiLSTM  &  \textbf{0.00896}   &  0.02533  & \textbf{0.93541} &  \textbf{0.12454} \\ 
                      & QNN  &  0.01324   &  \textbf{0.02096}   &  1.06966  & 0.36917 \\ \hline
\multirow{2}{*}{IAG.L}   & BiLSTM  &  0.02543   &  0.14314  & \textbf{0.87963} &  0.78823 \\ 
                      & QNN  &  \textbf{0.01527}  &  \textbf{0.06965}   &  1.02364  & \textbf{0.35888}  \\ \hline
\end{tabular}
\caption{The simulation results for both BiLSTM and QNN for all the signals (as shown in Figure~\ref{fig:results}). The bold entries indicate the best estimators for a given signal.}
\label{table:results}
\end{table}

\section{Summary and Discussion}
\label{sect:summ_discu}

We have compared the predictive power of QNNs, represented as a PQCs, and BiLSTM neural networks for a variety of temporal signals. Our comparative approach shows that PQCs can be used as a forecasting time series tool for stock price signals. In cases where the noise coefficient is small (noise amplitude noise variations being up to 40 per cent of the amplitude of the deterministic signal), then PQCs perform very similar to BiLSTM neural networks. However, whenever the noise coefficient is higher, the PQCs outperform them because the latter starts to slightly overfit, regardless of the optimiser, since it has 1830 times more parameters. 

Therefore, QNNs have two major advantages over traditional neural networks: (i) QNNs can be trained much faster than traditional neural networks (with a fully quantum optimisation algorithm) and (ii) QNNs can be more accurate, with smaller number of model parameters, for noisy signals with deterministic components. In our work, with respect to the training aspect, for the PQCs the forward propagation is quantum simulated, while the back propagation is purely classical. The entire training process of the PQCs can become fully quantum, once quantum continuous optimisation algorithms become more robust and widely tested. Research on quantum back propagation for some particular QNNs is already being conducted via the quantum approximate optimisation algorithm \cite{Verdon2019AQA} for optimising QNNs including variational quantum eigensolvers \cite{backprop}. Thus, in the future, if the optimisation process is replaced by a genuine backpropagation-like quantum optimisation algorithm then the network could be 
fully trained on a quantum computer.

Nevertheless, QNNs have limitations regarding to the volume of data that they can be trained upon. Currently, running complicated quantum machine learning models (i.e.~ with numerous model parameters) on \textit{big data} can be challenging even for quantum computers with large number of qubits such as IBM's Eagle chip with 127 qubits \cite{ball2021first}.

Moreover, in a realistic scenario, the entire end-to-end process for training a QNN, needs to be combined with a classical computer that would pre-process the data and prepare the input so it can be encoded to quantum arrays of qubits and after the readout qubit is measured, it would need to be scaled back to derive the prediction.

For Barclays, the application of time series forecasting using quantum machine learning is an active field of research including volatility prediction, asset prising and algorithmic trading. Future work could include the \textquotedblleft customer-centric\textquotedblright\; field of fraud prediction for transactional data using quantum anomaly detection algorithms such as variational quantum anomaly detection, inspired by autoencoders \cite{kottmann2021variational} and quantum generative adversarial networks \cite{Herr2021AnomalyDW}. This is a very challenging use case since in order to build a production-ready model one needs to train it using large data sets that cover at least one year of transactional activity (consisting of few billion transactions) in order to cover all seasonal trends. Also, due to the very low number of fraudulent transactions (unbalanced data) and the variety of fraudulent patterns the problem becomes even more complicated as creation of correct model features and precise sampling 
techniques need to occur during the data preprocessing phase.  

\section*{Acknowledgements}
We would like to thank Lee Braine (Barclays) for his helpful feedback.

\clearpage

\clearpage
\bibliographystyle{abbrv}
\bibliography{EmmanoulopoulosBarclaysQML}  

\end{document}